\documentclass[conference]{IEEEtran}
\IEEEoverridecommandlockouts
\usepackage{cite}
\usepackage{amsmath,amssymb,amsfonts}
\usepackage{algorithmic}
\usepackage{graphicx}
\usepackage{textcomp}
\usepackage{xcolor}
\usepackage{array}
\usepackage{tabularray}
\usepackage{hyperref}

\def\BibTeX{{\rm B\kern-.05em{\sc i\kern-.025em b}\kern-.08em
    T\kern-.1667em\lower.7ex\hbox{E}\kern-.125emX}}

\begin{document}

\title{Evaluating the Performance and Efficiency of Sentence-BERT for Code Comment Classification \\
%{\footnotesize \textsuperscript{*}Note: Sub-titles are not captured for https://ieeexplore.ieee.org  and should not be used}
}

\author{\IEEEauthorblockN{Fabian C. Peña}
\IEEEauthorblockA{
%\textit{dept. name of organization (of Aff.)} \\
\textit{University of Passau}\\
%Germany \\
fabiancamilo.penalozano@uni-passau.de}
\and
\IEEEauthorblockN{Steffen Herbold}
\IEEEauthorblockA{
%\textit{dept. name of organization (of Aff.)} \\
\textit{University of Passau}\\
%Germany \\
steffen.herbold@uni-passau.de}
}

\maketitle

\begin{abstract}
This work evaluates Sentence-BERT for a multi-label code comment classification task seeking to maximize the classification performance while controlling efficiency constraints during inference. Using a dataset of 13,216 labeled comment sentences, Sentence-BERT models are fine-tuned and combined with different classification heads to recognize comment types. While larger models outperform smaller ones in terms of F1, the latter offer outstanding efficiency, both in runtime and GFLOPS. As result, a balance between a reasonable F1 improvement (+0.0346) and a minimal efficiency degradation (+1.4x in runtime and +2.1x in GFLOPS) is reached.
\end{abstract}

\begin{IEEEkeywords}
Large Language Model (LLM), embeddings, code comment classification, efficiency evaluation.
\end{IEEEkeywords}

\section{Introduction}

Large Language Models (LLMs) have become popular tools to solve a broad range of Natural Processing Language (NLP) tasks. In the software engineering domain, LLMs can assist developers in writing code from free-text instructions, but also support other tasks across all the software life-cycle \cite{llm4se}. During the development stage, creating effective documentation is as important as writing functional and efficient code, enhancing readability and maintainability. It is considered a good practice that artifacts such as classes or methods include documentation in the form of comments describing aspects like purpose, responsibilities, parameters, among others.

The 4th International Workshop on NL-based Software Engineering (NLBSE 2025) has launched a competition inviting researchers to solve a code comment classification task \cite{nlbse2025}. The goal is to build a set of models which outperform the baselines in terms of classification performance while controlling efficiency constraints (runtime and GFLOPS) during inference. The task is multi-label and the provided dataset includes comments in three languages: Java, Python, and Pharo.

This paper documents the process of training and evaluating a set of models to solve the task in two stages: (1) generating vector representations of comments using embedding models, and (2) classifying these representations in at least one of the available labels. Candidate embedding models vary in terms of inference speed, size, and embedding dimension. For classification, different established algorithms are considered.

More details on the data set, competition rules, and baselines are found in \cite{nlbse2025} and Section \ref{dataset}. Sections \ref{approach} and \ref{results} detail the process followed to train and evaluate the candidate models. Finally, Section \ref{conclusions} highlights the most relevant findings and future opportunities. All the source code and detailed results can be found in the replication kit.\footnote{\url{https://github.com/aieng-lab/sbert-comment-classification/}}

\section{Dataset and Competition Rules}
\label{dataset}

The dataset provided for the competition is composed of comments extracted from 20 Java, Python, and Pharo open-source projects, manually labeled by experts. Labels represent comment types, i.e., semantic information contained in a fragment of the comment which follows a refined taxonomy based on actual commenting practices of developers \cite{rani2021, pascarella2017}. Comments are split into sentences using NEON \cite{disorbo2021}, resulting in 13,216 instances with at least one label. Finally, sentences are assigned to the train and test subsets with an approximate proportion of 80\%/20\% using stratified sampling. Table \ref{tbl-dataset} summarizes the composition of the dataset.

\begin{table}[t]
\caption{Code comments dataset overview}
\centering
\begin{tblr}{
  row{1} = {c},
  cell{2-4}{1-2} = {l, valign = m},
  cell{2-4}{3-4} = {c, valign = m},
  hlines,
}
\textbf{Language}   & \textbf{Labels}                                                                                                  & \textbf{Train} & \textbf{Test} \\
Java                & {Deprecation, expand, ownership,\\ pointer, rational, summary, usage}                                            & 7,614          & 1,725         \\
Python              & {Development notes, expand\\ parameters, summary, usage}                                                         & 1,884          & 406           \\
Pharo               & {Class references, collaborators\\ example, intent, key implementation\\ points, key messages, responsibilities} & 1,298          & 289           
\end{tblr}
\label{tbl-dataset}
\end{table}

While some basic text preprocessing is already applied (e.g., lower-casing, removal of special characters, normalization), some sentences still contain comment styling characters like \texttt{/**}, \texttt{//}, \texttt{*/}, and \texttt{\#} in the Java and Python subsets. Additionally, some sentences are extremely short in terms of number of tokens. Moreover, the most common labels include \textit{example}, \textit{parameters}, and \textit{summary} while others such as \textit{class references}, \textit{deprecation}, and \textit{ownership} are highly imbalanced. Although the task is defined as multi-label, just ~4\% of the sentences have more than one label.

%The 10th lower percentile have less than 5, 3, and 2 tokens in the Java, Python, and Pharo subsets, respectively. In the worst cases, sentences have a single character.

%\section{Competition Rules}
%\label{rules}

The goal of the competition is to build a set of models (one per each language) to classify sentences in at least one label. Although it is important to have high classification performance in the test subsets, submitted models are expected to run efficiently, i.e., to make inferences in a reasonable runtime requiring a limited number of GFLOPS. The formula to rank competitors, aggregated for the three languages, is:
\begin{multline}
submission\_score = 0.6 \times avg\_F_1 \\
 + 0.2 \times \frac{(5 \, \text{secs.} - avg\_runtime)}{5 \, \text{secs.})} \\
 + 0.2 \times \frac{(5,000 \, \text{GFLOPS} - avg\_GFLOPS)}{5,000 \, \text{GFLOPS})}
 \label{eq-submission-score}
\end{multline}

Note that an increase in $6.\bar{6}$ in F1, a decrease of 1 sec. in runtime or 1,000 GFLOPS are equivalent. This is the reason why optimizing for runtime and model size is important, as changes in model efficiency have a large impact. Furthermore, $submission\_score$ aggregates the results of the three languages, meaning that optimizing the runtime and GFLOPS required for Java is around 4x and 6x more important than for Python and Pharo, respectively, due to the different sizes of the test subsets. Similarly, since there are seven different labels for Java and Pharo, and only five for Python, Java and Pharo errors affect the average F1 more than Python.

%Therefore, results per language are also reported to enable analysis without these biasing factors.

%\section{Baselines}
%\label{baselines}

Because model efficiency is conditioned by hardware acceleration, final evaluation of the submitted models must be carried out in Google Colab T4. 

The competition also provides three baseline models. Following some of the ideas presented in the literature \cite{alkaswan2023stacc}, these models are built using SetFit \cite{setfit}. This framework extends Sentence-BERT \cite{sbert}, enabling developers to easily fine-tune sentence embedding models for classification tasks. The baselines use the pre-trained model \texttt{paraphrase-MiniLM-L3-v2} with a multi-output logistic regression head. The official $submission\_score$ reported for these models, calculated in Google Colab T4, is 0.6968.

\section{Approach}
\label{approach}

While consolidated LLMs like BERT \cite{bert} are widely spread and have strong support from the community, they are optimized to predict tokens, not to provide suitable vector representations (i.e., embeddings) of sentences. Sentence-BERT \cite{sbert} solves this problem by learning to estimate the similarity between pairs of sentences. Additionally, these kind of models tend to be smaller by up to an order of magnitude representing an advantage for the competition due to the efficiency constraints. Table \ref{tbl-sentence-models} lists the pre-trained embedding models considered for the experimentation. These models are selected as potential good candidates due to their inference speed, size and performance previously reported on diverse sentence embedding tasks.\footnote{\url{http://www.sbert.net/docs/sentence\_transformer/pretrained\_models.html}}

For the first experimentation stage, these embedding models are fine-tuned using contrastative learning on different portions of the training subsets to enable the generation of meaningful domain-specific sentence representations, and complemented with a multi-output logistic regression head by default. Under the hypothesis that using a larger number of positive and negative pairs helps to produce better representations and consequently higher classification performance, the parameter \texttt{num\_iterations} varies from 20 to 60, where the number of positive and negative pairs is calculated as $num\_iterations \times num\_sentences \times 2$. The evaluation considers the runtime and GFLOPS required by each embedding model and their contribution to the $submission\_score$.

\begin{table}[t]
\caption{Pre-Trained embedding models for experimentation}
\centering
\begin{tblr}{
  row{1} = {c, valign=m},
  cell{2}{2} = {c},
  cell{2}{3} = {c},
  cell{2}{4} = {c},
  cell{3}{2} = {c},
  cell{3}{3} = {c},
  cell{3}{4} = {c},
  cell{4}{2} = {c},
  cell{4}{3} = {c},
  cell{4}{4} = {c},
  cell{5}{2} = {c},
  cell{5}{3} = {c},
  cell{5}{4} = {c},
  cell{6}{2} = {c},
  cell{6}{3} = {c},
  cell{6}{4} = {c},
  hline{1-2,7} = {-}{},
}
\textbf{Embedding model}   & {\textbf{Speed}\\\textbf{(sent./sec.)}} & {\textbf{Size}\\\textbf{(MB)}} & {\textbf{Embedding}\\\textbf{dimensions}} \\
paraphrase-MiniLM-L3-v2    & 19,000                                    & 61                               & 384                                         \\
all-MiniLM-L6-v2           & 14,200                                    & 80                               & 384                                         \\
paraphrase-albert-small-v2 & 5,000                                     & 43                               & 768                                         \\
all-distilroberta-v1       & 4,000                                     & 290                              & 768                                         \\
all-mpnet-base-v2          & 2,800                                     & 420                              & 768                                         
\end{tblr}
\label{tbl-sentence-models}
\end{table}

Effective classification depends on both meaningful sentence representations and careful optimization of the classification head, which directly impacts task performance. For the second experimentation stage, the previously fine-tuned embedding models are combined with more robust classification heads such as random forest (RF), xgboost (XG), and support vector machines (SVM). During the process of training the classification head candidates, some key hyper-parameter values are considered. For RF and XG, the most important hyper-parameter to experiment with is \texttt{max\_depth} with values ranging from 3 to 20. For SVM, typical values for \texttt{kernel} include \textit{linear}, \textit{poly}, \textit{rbf}, and \textit{sigmoid}. For both SVM and LR, it is a good practice to apply regularization by setting \texttt{C} between 0.001 to 1.0 with exponential step sizes.%\footnote{This regularization is also how this logistic regression is different from the usually used logistic regression through softmax activation as final layer in neural networks.}

%In order to speed up all the training and evaluation process of candidates, a local environment including 8 GPUs NVDIA A100-SXM4-80GB is used. Most of the results reported in this work are obtained using this environment, however, the final evaluation of the submitted models is also carried out in Google Colab T4. Each scenario regarding when one or another environment has been used is appropriately clarified.

The training and evaluation process leverages an 8-GPU NVIDIA A100 setup. However, final evaluation of the submitted models is also carried out in Google Colab T4, as specified in the competition rules.

\section{Results}
\label{results}

\begin{figure}[t]
\centerline{\includegraphics[width=0.8\columnwidth]{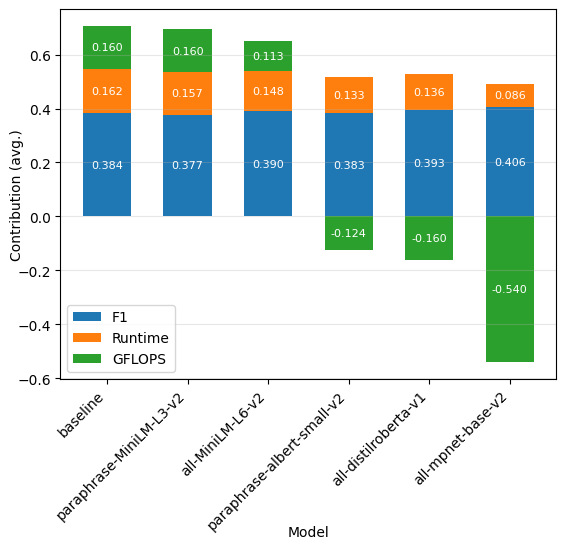}}
\caption{Contributions to the $submission\_score$ by embedding model.}
\label{fig-contributions}
\end{figure}

Table \ref{tbl-results-1} summarizes the results of the first experimentation stage obtained in the local setup. The largest embedding model, \texttt{all-mpnet-base-v2}, yields the highest classification performance with an improvement of +0.0527 in average F1 compared to the baselines. However, due to slow runtime and larger computational demands (i.e., GFLOPS), this model actually achieves the worst $submission\_score$ among all the candidates. This effect is better explained in Figure \ref{fig-contributions}, where positive and negative contributions of the three evaluation aspects are represented. Even with a F1 of 1.0, the penalties incurred due to the GFLOPS required by \texttt{all-mpnet-base-v2} would make it impossible to outperform the baseline models, which remain as the best candidates in terms of $submission\_score$. Note that the $submission\_score$ for the baseline models, calculated in the local setup, does not differ much from the official one presented at the end of Section \ref{dataset}.

\begin{table*}[ht]
\caption{Results of Experimenting with Different Pre-Trained Embedding Models}
\centering
\begin{tblr}{
  row{1} = {c},
  column{even} = {c},
  column{3} = {c},
  column{5} = {c},
  cell{3}{5} = {r=3}{},
  cell{6}{5} = {r=3}{},
  cell{9}{5} = {r=3}{},
  cell{12}{5} = {r=3}{},
  cell{15}{3} = {c},
  cell{15}{5} = {r=3}{},
  hline{1-3,6,9,12,15,18} = {-}{},
}
\textbf{Embedding model}   & \textbf{Iterations} & \textbf{Avg. F1} & \textbf{Avg. runtime} & \textbf{Avg. GFLOPS} & \textbf{Submission score} \\
baselines                   & 20                  & 0.6394           & 0.9422                & 999.0271             & 0.7060                    \\
paraphrase-MiniLM-L3-v2    & 20                  & 0.6348           & 1.0916                & 999.0271             & 0.6973                    \\
paraphrase-MiniLM-L3-v2    & 40                  & 0.6246           & 1.0077                &                      & 0.6945                    \\
paraphrase-MiniLM-L3-v2    & 60                  & 0.6243           & 1.0931                &                      & 0.6909                    \\
all-MiniLM-L6-v2           & 20                  & 0.6425           & 1.3468                & 2173.2976            & 0.6447                    \\
all-MiniLM-L6-v2           & 40                  & 0.6476           & 1.2921                &                      & 0.6499                    \\
all-MiniLM-L6-v2           & 60                  & 0.6578           & 1.2674                &                      & 0.6571                    \\
paraphrase-albert-small-v2 & 20                  & 0.6363           & 1.5817                & 8088.5031            & 0.3950                    \\
paraphrase-albert-small-v2 & 40                  & 0.6342           & 1.5483                &                      & 0.3950                    \\
paraphrase-albert-small-v2 & 60                  & 0.6442           & 1.8769                &                      & 0.3879                    \\
all-distilroberta-v1       & 20                  & 0.6499           & 1.5599                & 8997.9049            & 0.3676                    \\
all-distilroberta-v1       & 40                  & 0.6559           & 1.7063                &                      & 0.3654                    \\
all-distilroberta-v1       & 60                  & 0.6585           & 1.5653                &                      & 0.3725                    \\
all-mpnet-base-v2          & 20                  & 0.6921           & 3.5339                & 18489.3073           & -0.0657                   \\
all-mpnet-base-v2          & 40                  & 0.6795           & 2.6626                &                      & -0.0384                   \\
all-mpnet-base-v2          & 60                  & 0.6564           & 2.3183                &                      & -0.0384                   
\end{tblr}
\label{tbl-results-1}
\end{table*}

For the second stage, Table \ref{tbl-results-2} shows the model type and hyper-parameters corresponding to the best classification head for each embedding model, calculated in the local setup. As observed, all the candidates have opportunity for improvement in terms of classification performance when using more robust classification heads, although at a cost in efficiency. Still, for Python and Pharo, top-ranked candidates are slightly slower and larger than the baseline models compared to most of their counterparts, yielding a superior $submission\_score$ of +0.0244 for Python and +0.0156 for Pharo.

When evaluating the best candidates with non-default head across languages in the local setup, the $submission\_score$ is 0.6709. However, for Java, note that the combination of baseline embedding model and default head is still the best in terms of $submission\_score$. When using strictly the best candidates for the three languages (i.e., baseline embedding model and default head for Java), a improvement of +0.0347 in the $submission\_score$ is evidenced. This slightly worse $submission\_score$ for the models to be submitted, i.e., the best candidates with non-default head, is mostly explained by the embedding model \textit{all-MiniLM-L6-v2} which is slower and larger than \textit{paraphrase-MiniLM-L3-v2}, same pre-trained embedding model used for the baselines. Additionally, when running the evaluation of the models to be submitted in Google Colab T4, the $submission\_score$ is 0.6536, where this reduction is explained by the degradation in the conditions of the acceleration hardware.\footnote{\url{https://colab.research.google.com/drive/17Bep6v\_1Ia\_dVKNnVtg\_myr7GMRhPfn1?usp=sharing}}

Finally, because most of the sentences only have one label, it would be possible to use a simple bag-of-words with multinomial Naive Bayes as classifier (single class) and achieve a competitive $submission\_score$ of 0.6828. Since the models are small, the runtime is only 0.0360 secs. and basically no GFLOPS are required. However, an average F1 of 0.4736 raises concerns about the practical viability of this approach.

\begin{table*}[ht]
\caption{Results of Experimenting with Different Classification Heads.}
\centering
\begin{tblr}{
   row{23} = {c},
  cell{1}{1} = {c},
  cell{1}{2} = {c},
  cell{1}{3} = {c},
  cell{1}{5} = {c},
  cell{1}{6} = {c},
  cell{1}{7} = {c},
  cell{2}{1} = {r=7}{},
  cell{2}{4} = {c},
  cell{2}{5} = {c},
  cell{2}{6} = {c},
  cell{2}{7} = {c},
  cell{3}{4} = {c},
  cell{3}{5} = {c},
  cell{3}{6} = {c},
  cell{3}{7} = {c},
  cell{4}{4} = {c},
  cell{4}{5} = {c},
  cell{4}{6} = {c},
  cell{4}{7} = {c},
  cell{5}{4} = {c},
  cell{5}{5} = {c},
  cell{5}{6} = {c},
  cell{5}{7} = {c},
  cell{6}{4} = {c},
  cell{6}{5} = {c},
  cell{6}{6} = {c},
  cell{6}{7} = {green,c},
  cell{7}{4} = {c},
  cell{7}{5} = {c},
  cell{7}{6} = {c},
  cell{7}{7} = {c},
  cell{8}{4} = {c},
  cell{8}{5} = {c},
  cell{8}{6} = {c},
  cell{8}{7} = {c},
  cell{9}{1} = {r=7}{},
  cell{9}{4} = {c},
  cell{9}{5} = {c},
  cell{9}{6} = {c},
  cell{9}{7} = {c},
  cell{10}{4} = {c},
  cell{10}{5} = {c},
  cell{10}{6} = {c},
  cell{10}{7} = {c},
  cell{11}{4} = {c},
  cell{11}{5} = {c},
  cell{11}{6} = {c},
  cell{11}{7} = {green,c},
  cell{12}{4} = {c},
  cell{12}{5} = {c},
  cell{12}{6} = {c},
  cell{12}{7} = {c},
  cell{13}{4} = {c},
  cell{13}{5} = {c},
  cell{13}{6} = {c},
  cell{13}{7} = {c},
  cell{14}{4} = {c},
  cell{14}{5} = {c},
  cell{14}{6} = {c},
  cell{14}{7} = {c},
  cell{15}{4} = {c},
  cell{15}{5} = {c},
  cell{15}{6} = {c},
  cell{15}{7} = {c},
  cell{16}{1} = {r=7}{},
  cell{16}{4} = {c},
  cell{16}{5} = {c},
  cell{16}{6} = {c},
  cell{16}{7} = {c},
  cell{17}{4} = {c},
  cell{17}{5} = {c},
  cell{17}{6} = {c},
  cell{17}{7} = {c},
  cell{18}{4} = {c},
  cell{18}{5} = {c},
  cell{18}{6} = {c},
  cell{18}{7} = {c},
  cell{19}{4} = {c},
  cell{19}{5} = {c},
  cell{19}{6} = {c},
  cell{19}{7} = {c},
  cell{20}{4} = {c},
  cell{20}{5} = {c},
  cell{20}{6} = {c},
  cell{20}{7} = {c},
  cell{21}{4} = {c},
  cell{21}{5} = {c},
  cell{21}{6} = {c},
  cell{21}{7} = {c},
  cell{22}{4} = {c},
  cell{22}{5} = {c},
  cell{22}{6} = {c},
  cell{22}{7} = {green,c},
  cell{23}{1} = {c=3}{},
  cell{23}{7} = {green,c},
  cell{24}{1} = {c=3}{},
  cell{24}{4} = {c},
  cell{24}{5} = {c},
  cell{24}{6} = {c},
  cell{24}{7} = {c},
  cell{25}{1} = {c=3}{},
  cell{25}{4} = {c},
  cell{25}{5} = {c},
  cell{25}{6} = {c},
  cell{25}{7} = {c},
  cell{26}{1} = {c=3}{},
  cell{26}{4} = {c},
  cell{26}{5} = {c},
  cell{26}{6} = {c},
  cell{26}{7} = {c},
  cell{27}{1} = {c=3}{},
  cell{27}{4} = {c},
  cell{27}{5} = {c},
  cell{27}{6} = {c},
  cell{27}{7} = {green,c},
  cell{28}{1} = {c=3}{},
  cell{28}{4} = {c},
  cell{28}{5} = {c},
  cell{28}{6} = {c},
  cell{28}{7} = {c},
  hline{1-2,9,16,23,27} = {-}{},
  hline{3,10,17} = {2-7}{},
  hline{23} = {2}{-}{},
  hline{27} = {2}{-}{},
  hline{29} = {2}{-}{},
}
\textbf{Language}                                                               & \textbf{ Embedding model}  & \textbf{Head}                & \textbf{Avg. F1} & \textbf{Avg. runtime} & \textbf{Avg. GFLOPS} & \textbf{Submission score} \\
Java                                                                            & baseline                   & default                      & 0.6979           & 0.6750                & 803.4690             & 0.7596                    \\
                                                                                & all-mpnet-base-v2          & SVM, C: 0.01, kernel: rbf    & 0.7404           & 3.7043                & 15342.3384           & 0.0824                    \\
                                                                                & all-distilroberta-v1       & SVM, C: 0.1, kernel: linear  & 0.7378           & 2.6844                & 7527.8992            & 0.4342                    \\
                                                                                & paraphrase-albert-small-v2 & SVM, C: 0.01, kernel: rbf    & 0.7277           & 3.2573                & 6375.5939            & 0.4513                    \\
                                                                                & all-MiniLM-L6-v2           & RF, max\_depth: 9            & 0.7251           & 0.9127                & 1782.4005            & 0.7273                    \\
                                                                                & paraphrase-MiniLM-L3-v2    & LR, C: 0.01                  & 0.7119           & 2.1144                & 803.4690             & 0.7104                    \\
                                                                                & baseline                   & SVM, C: 0.01, kernel: linear & 0.7022           & 2.1690                & 803.4690             & 0.7024                    \\
Python                                                                          & baseline                   & default                      & 0.6030           & 0.2351                & 103.6213             & 0.7483                    \\
                                                                                & all-distilroberta-v1       & RF, max\_depth: 6            & 0.6737           & 0.3447                & 811.9539             & 0.7580                    \\
                                                                                & all-MiniLM-L6-v2           & RF, max\_depth: 4            & 0.6564           & 0.3212                & 207.1154             & 0.7727                    \\
                                                                                & all-mpnet-base-v2          & XG, max\_depth: 2            & 0.6548           & 0.3648                & 1665.6587            & 0.7117                    \\
                                                                                & paraphrase-MiniLM-L3-v2    & LR, C: 0.1                   & 0.6165           & 1.5583                & 103.6213             & 0.7034                    \\
                                                                                & paraphrase-albert-small-v2 & LR, C: 0.01                  & 0.6157           & 1.8046                & 966.8189             & 0.6586                    \\
                                                                                & baseline                   & LR, C: 0.01                  & 0.6062           & 1.5670                & 103.6213             & 0.6969                    \\
Pharo                                                                           & baseline                   & default                      & 0.6068           & 0.1973                & 91.9368              & 0.7525                    \\
                                                                                & all-mpnet-base-v2          & default                      & 0.7105           & 0.3296                & 1481.3102            & 0.7539                    \\
                                                                                & all-distilroberta-v1       & RF, max\_depth: 6            & 0.6658           & 0.3014                & 658.0518             & 0.7611                    \\
                                                                                & paraphrase-albert-small-v2 & SVM, C: 1.0, kernel: sigmoid & 0.6579           & 1.6541                & 746.0903             & 0.6988                    \\
                                                                                & paraphrase-MiniLM-L3-v2    & SVM, C: 1.0, kernel: poly    & 0.6414           & 1.5860                & 91.9368              & 0.7177                    \\
                                                                                & all-MiniLM-L6-v2           & RF, max\_depth: 6            & 0.6365           & 0.2838                & 183.7817             & 0.7632                    \\
                                                                                & baseline                   & RF, max\_depth: 5            & 0.6356           & 0.2394                & 91.9368              & 0.7681                    \\
\textit{Cross language results combining the best candidates with non-default head} &                            &                              & 0.6740           & 1.2567                & 2081.4527            & 0.6709 \\
\textit{Cross language results combining the best candidates (default head for Java)} &                            &                              & 0.6640           & 1.2165                & 1102.5212            & 0.7056 \\
\textit{Cross language results combining the baselines with default heads} & & & 0.6394 & 0.9422 & 999.0271 & 0.7060 \\
\textit{Cross language results using multinomial Naive Bayes with bag-of-words} & & & 0.4736 & 0.0360 & $0^*$ & 0.6828 \\
\textit{Final results for submitted models, calculated in Google Colab T4} &                            &                              & 0.6740           & 1.6859                & 2084.5110            & 0.6536 \\
\textit{Official results for baseline models, calculated in Google Colab T4} &                            &                              & 0.6394           & 1.1702                & 999.0271            & 0.6968 \\
\end{tblr}
\label{tbl-results-2}
\end{table*}

\section{Conclusions}
\label{conclusions}

This work focuses on training and evaluating models that yield superior classification performance over the baselines for the NLBSE 2025 competition. The results show that improvements in F1 are relatively straightforward to achieve, both with different embedding models (Stage 1) as well as with different classification heads (Stage 2). However, all of these improvements often come with efficiency degradations. Since these degradations are strongly penalized within the competition, it is actually difficult to achieve consistent improvements in $submission\_score$ using the current approach.

An alternative would be not to focus mostly on model training, but rather on aspects that try to make models smaller without large losses to the classification performance, e.g., using quantization or distillation techniques. Considering the strong focus of $submission\_score$ on runtime and GFLOPS, even a small increase in these aspects could counteract losses in F1. Taken to the extreme, small models like a simple Naive Bayes also work, though in practice this approach is not suggested as the F1 is extremely low.

\section*{Acknowledgments}

This work is funded by the German Research Foundation (DFG) through the project SENLP, grant 524228075.

\bibliographystyle{IEEEtran}
\bibliography{refs}

\end{document}